\documentstyle[11pt,dunk2001_asp,twoside, psfig]{article}
\def\edcomment#1{\iffalse\marginpar{\raggedright\sl#1\/}\else\relax\fi}
\reversemarginpar

\begin{document}
\title{Chemical Abundance Variations in Globular Clusters: Recent Results 
	from Mildly Metal-Poor M5}
 \author{Inese I.~Ivans}
\affil{Astronomy Dept.~\& McDonald Observatory, Univ.~of Texas, Austin, TX, 
	USA/Research School of Astronomy and Astrophysics, Australian 
	National Univ., Canberra, A.C.T., Australia}
\author{Robert P.~Kraft}
\affil{UCO-Lick Observatory, Univ.~of California, Santa Cruz, CA, USA}
\author{Chris Sneden}
\affil{Astronomy Dept.~\& McDonald Obs., Univ.~of Texas, Austin, TX, USA}
\author{Graeme H.~Smith}
\affil{UCO-Lick Observatory, Univ.~of California, Santa Cruz, CA, USA}
\author{R.~Michael Rich}
\affil{Dept.~of Physics \& Astronomy, UCLA, Los Angeles, CA, USA}
\author{Matthew Shetrone}
\affil{McDonald Obs.~\& Univ.~of Texas at Austin, Fort Davis, TX, USA}

\begin{abstract}
We present a chemical composition analysis of 36 giant stars in mildly 
metal-poor globular cluster M5.  In comparing the M5 results to those 
obtained in M4, a cluster previously considered to be a ``twin'' in age, 
metallicity and chemical composition, we find large star-to-star 
variations in the abundances of elements sensitive to proton-capture 
nucleosynthesis, similar [Fe/H] values, but factor of two differences in 
some $\alpha$-capture, odd-Z and slow neutron-capture process elements.  
Among stars in globular clusters, apparently there are no definitive 
``single'' values of [el/Fe] at a given [Fe/H] for many important 
elements.
\end{abstract}

\section{Introduction}
Large star-to-star abundance variations in C, N, O, Na, Mg and Al exist
among bright giant stars in metal-poor globular clusters.  Star-to-star 
abundance variations have been found in all metal-poor globular clusters 
in which the variations have been sought.  In clusters with sufficiently 
large sample sizes, N is typically anti-correlated with O and C, Na is 
anti-correlated with O, and Al is anti-correlated with Mg.  The reader 
is referred to \S1 of Ivans et al.~(2001) for references to recent 
relevant reviews.

The abundance anti-correlations found among cluster stars likely result 
from proton-capture nucleosynthesis (where C and N are converted into N, 
Ne into Na, and Mg into Al).  What is less clear is whether the synthesis 
takes place in the giants we presently observe (evolutionary scenario) or 
in a prior generation of more massive evolved stars that polluted the gas 
from which the present generation of stars was formed (primordial 
scenario).  One expectation of the evolutionary scenario is that the 
distribution of these abundances should change with advancing evolutionary 
state.  Thus as evolution proceeds, one might expect to find relatively 
more stars with low O and Mg and fewer with high O and Mg, and 
correspondingly more with high Na and Al and fewer with low Na and Al.  

Given the importance of decoupling the evolutionary effects from primordial
enrichments, we performed an abundance study of a sample of 36 giant 
stars in the mildly metal-poor globular cluster M5, to compare against the 
Ivans et al.~(1999) large-sample study of the similar metallicity cluster 
M4.  Details of the observations and analysis techniques are to be found in 
Ivans et al.~(2001). 

\section{Abundance Results}
Our M5 stars display the ``classic'' anti-correlations and correlations of 
O, Na and Al, the elements that are sensitive to proton-capture 
nucleosynthesis, as previously observed in brighter M5 stars as well as in 
other clusters observed by the Lick-Texas group (see Figure 1).  These 
abundance patterns also correlate with the CN strengths.  When the RGB + 
tip stars are binned into two evolutionary groups by log~g, the groups 
possess statistically significant different means of distribution in 
[O/Fe].  On average, stars with lower log~g values have higher O and lower 
Na abundances than stars with higher log~g values.  

\begin{figure}[bht]
\centerline{\vbox{
\psfig{figure=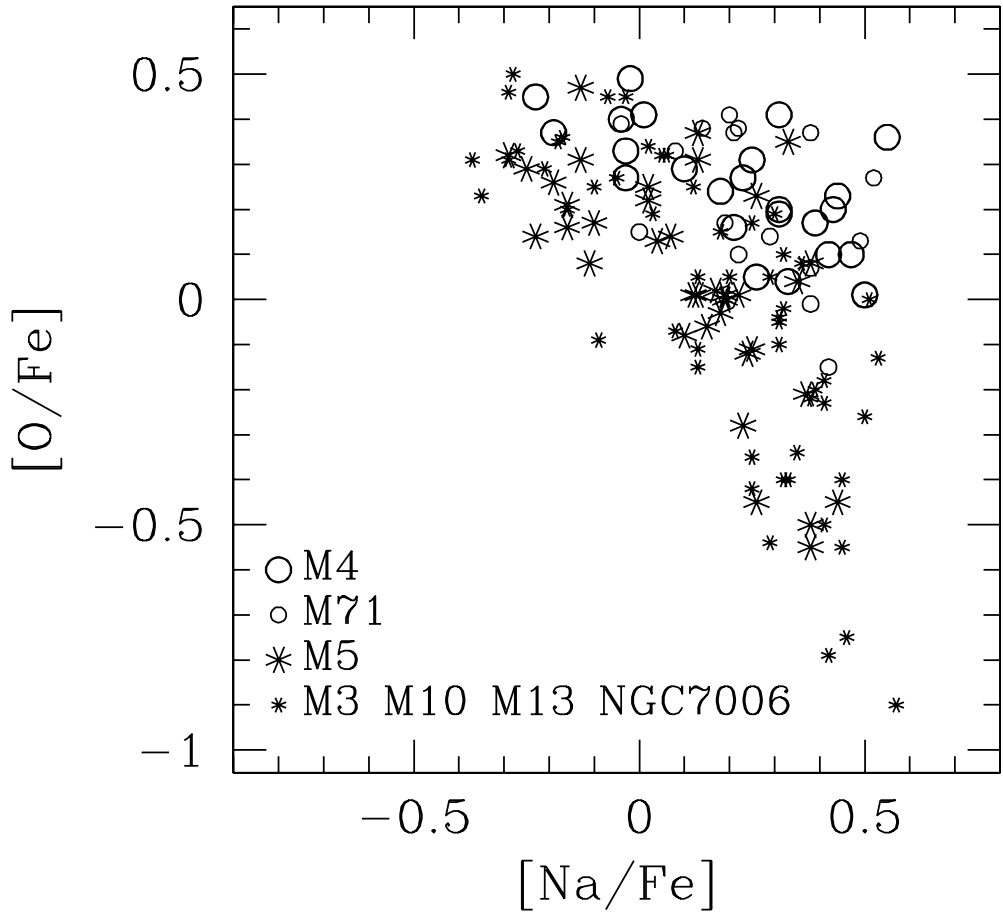}
}}
\caption{[Na/Fe] vs [O/Fe] for M5 and M4 in the context of globular 
clusters with similar metallicities previously studied by the Lick-Texas 
group (adapted from Fig.~16, Ivans et al.~2001).}
\end{figure}

Between clusters, we find that the variations of O and Na are correlated 
but the degree of variation appears to differ.  The O vs Na 
anti-correlation in M5 resembles that found in the more metal-poor 
clusters M3, M10, M13 and NGC7006 ([Fe/H] $\sim$ --1.5 to --1.6).  M4's 
behaviour seems to be more like that of M71, a disk cluster of higher 
metallicity ([Fe/H] $\sim$ --0.7).  The range of these abundance variations 
seems to correlate as well with other observables.  For instance, for these
7 clusters, in addition to binning by the ``first parameter'' (metallicity), 
the clusters can be binned by their ``second parameter'' (as quantified by 
the horizontal branch ratio) as well as by their orbital inclination.

In M5, we find that the abundance ratios for Mg, Si, Ca, Sc, Ti, V, Ni, Ba 
and Eu show no significant abundance variations and that these ratios are 
comparable to those of halo field stars of similar metallicities.  However, 
in comparing the abundances of M5 to M4, we find that Si, Al, Ba and La are 
overabundant in M4 with respect to what is seen in M5, confirming and 
expanding results from previous studies (Brown \& Wallerstein 1992; Ivans 
et al.~1999).  In Figure 2, we display boxplots of these four elemental 
abundances as well as Eu, an $r$-process element, for M5, M4 and halo field 
stars of comparable metallicities.  The ``box'' in each case contains the 
middle 50\% 
of the data and a horizontal line indicates the median value of a 
particular element.  The vertical tails indicate the total range of 
abundances, excluding mild outliers which are denoted by open circles.  


\begin{figure}[bht]
\centerline{\vbox{
\psfig{figure=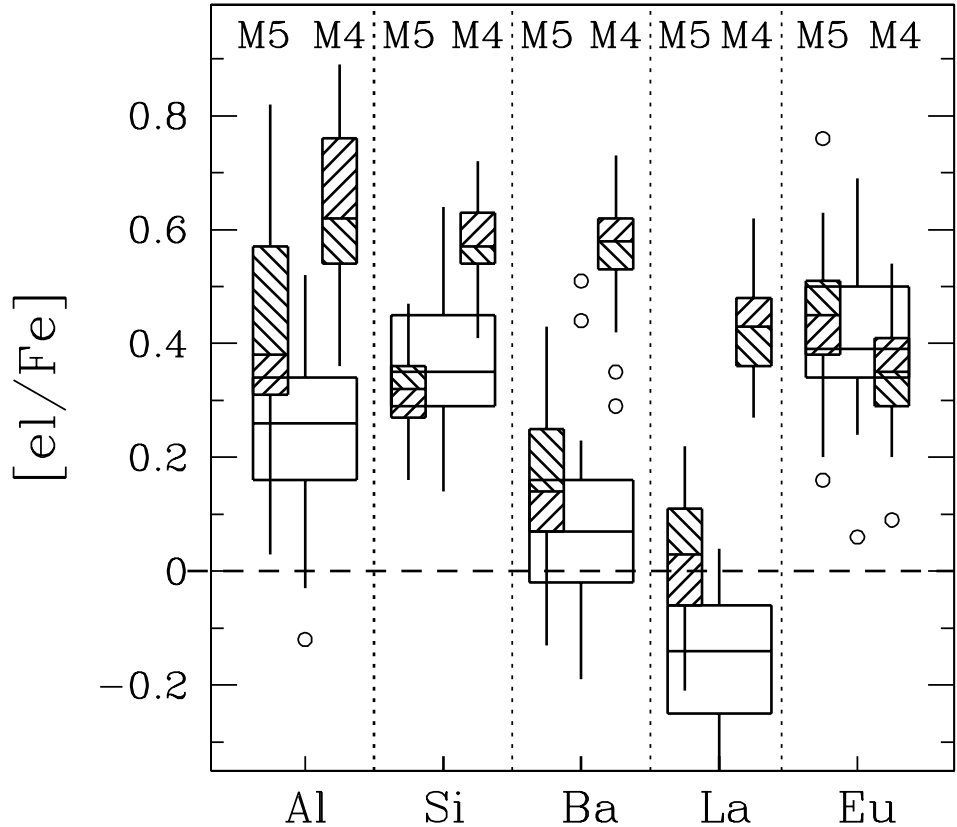}
}}
\caption{Boxplots of abundances of some odd-Z, $\alpha$-capture,  
$s$- and $r$-process elements. In each panel, we show shaded boxplots 
for M5 (on the left) and M4 (on the right), both superimposed on an
unshaded boxplot derived for the field halo stars of comparable 
metallicity.
}
\end{figure}

\section{Summary and Conclusions}
In M5, we find correlations between CN, O, Na and Al that are both 
consistent with those seen in previous globular cluster studies and that
follow the expected pattern of proton-capture nucleosynthesis.  With 
clusters that bracket M4 and M5 in metallicity, we find that the abundance 
patterns can be divided into two groups: the O vs Na anti-correlation 
found in M5 resembles that found in slightly more metal-poor globular 
clusters M3, M10 and M13 whereas the M4 pattern resembles that of the 
more metal-rich disk cluster M71. The cluster similarities extend to the
horizontal branch morphologies.  We find good agreement between M5, M4 
and field stars of comparable metallicity in the Fe-peak and 
$\alpha$-element abundances, with the exception of a Si overabundance in 
M4.   Ba and La are similarly overabundant in M4 with respect to M5 and 
the field, as is Al.  Based on these large stellar samples for M5 and M5, 
we extend previous findings and conclude that there is no ``single'' 
value of [el/Fe] at a given [Fe/H] for at least some $\alpha$-capture, 
odd-Z and $s$-process elements, in this case Si, Al, Ba and La.

\acknowledgments
We are happy to acknowledge that this research was funded by US NSF grants 
AST-9618351 to R.P.K.~and G.H.S.~and AST-9618364 and AST-9987162 to C.S.  
I.I.I.~also sincerely thanks the meeting organizers for their support of my 
attendance at this excellent Workshop and the Australian Federation 
of University Women (Queensland) for the Audrey Jorss Commemorative 
Fellowship financial support during the time that this work was performed.

\end{document}